\documentstyle[11pt,mrs2001,epsfig,indentfirst]{article}
\begin{document}
\title{2-POINT CORRELATION FUNCTION OF HI-SELECTED GALAXIES}

\author{S. TANTISRISUK$^1$, R. WEBSTER$^1$}
\affil{$^1$School of Physics, University of Melbourne, Parkville, AUSTRALIA 3010}

\begin{abstract}
The 2-point spatial correlation function (CF), $\xi(s)$, has
been used to study the clustering of the galaxies in the preliminary
version of the HIPASS Bright
Galaxy Catalogue (HIPASS BGC), which includes the 1,000 HI brightest 
galaxies in the Southern sky. This is the first time the CF has been used to analyse an HI-selected sample. The CF is well described by a power law,
$\xi(s)\sim(s/s_0)^{-\gamma}$, with slope $\gamma\sim1.7$ and
correlation length $s_0\sim3.55$ $h^{-1}$ Mpc using Peebles estimator. However,
when the Hamilton estimator is used, the CF is fit with $\gamma\sim2.19$
and $s_0\sim3.37$ $h^{-1}$ Mpc. Note that, these HI-selected galaxies show less
clustering than optically-selected surveys which have a correlation length $s_0\sim$ 5.5  $h^{-1}$ Mpc  or larger.\\
\end{abstract}

\section{Introduction}
The CF is one of the most popular statistical tools used to quantify the degree of galaxy clustering. It also has been used to characterize the dependence of the galaxy clustering on the properties of galaxies for example morphology \cite{myref2}, surface brightness, luminosity and internal dynamics.\\
\indent Generally, the CF is fit within the range $\gamma\sim$ 1.5-2.3 and $s_0\sim$ 5-7.5 $h^{-1}$ Mpc for optically-selected galaxies, where $H_0$ = 100 $h$ km s$^{-1}$ Mpc$^{-1}$. The data being used for analyzing the CF here is HIPASS BGC [6], which includes the 1,000 HI brightest galaxies from HI Parkes All Sky Survey (HIPASS; [1]). The survey is the largest, least-biased survey of large-scale structure in HI-selected galaxies. Since the survey has been done in radio regime so it can easily image through the Milky Way plane unlike optical or infrared surveys. Approximately 70\% of the new galaxies have been found in the Zone of Avoidance $|b| < 10 \deg$.\\
%\indent In section 2, the HIPASS BGC is briefly described. In section 3, the results of CF from HIPASS BGC have been shown and discussed. Conclusion is drawn in section 4.

\section{HIPASS BGC}
The HIPASS survey has imaged the entire Southern sky and 40\% of Northern sky using the Parkes multibeam receiver. The survey has been looking for galactic and extragalactic neutral hydrogen (HI) and is sensitive to a volumn of six million cubic Mpc. This survey is expected to eventually cataloque about 10,000 galaxies. The HIPASS BGC  includes the 1,000 HI-brightest galaxies ($\delta < 2\deg$) according to their HI peak flux in the global HI spectrum (peak$\geq$ 117 mJy/beam) with a systemic velocity 350 - 8,000 km/s. All known galaxies with systemic velocity $<$ 350 km/s have been added in the sample. Figure 1a shows distribution of HIPASS BGC as a function distance, with the selection function multiplied by area, $s^2\xi(s)$. The peak of distribution is around 13 $h^{-1}$ Mpc.
\section{The CF in HIPASS BGC}
Two estimators have been used for analyzing the CF in HIPASS BGC:
\begin{equation}
\xi_{\textup{DP}}=\frac{DD}{DR}-1,{\textup{ }} \xi_{\textup{Ham}}=\frac{DD . RR}{{DR}^2}-1,\\
\end{equation}
\begin{figure}
\plottwo{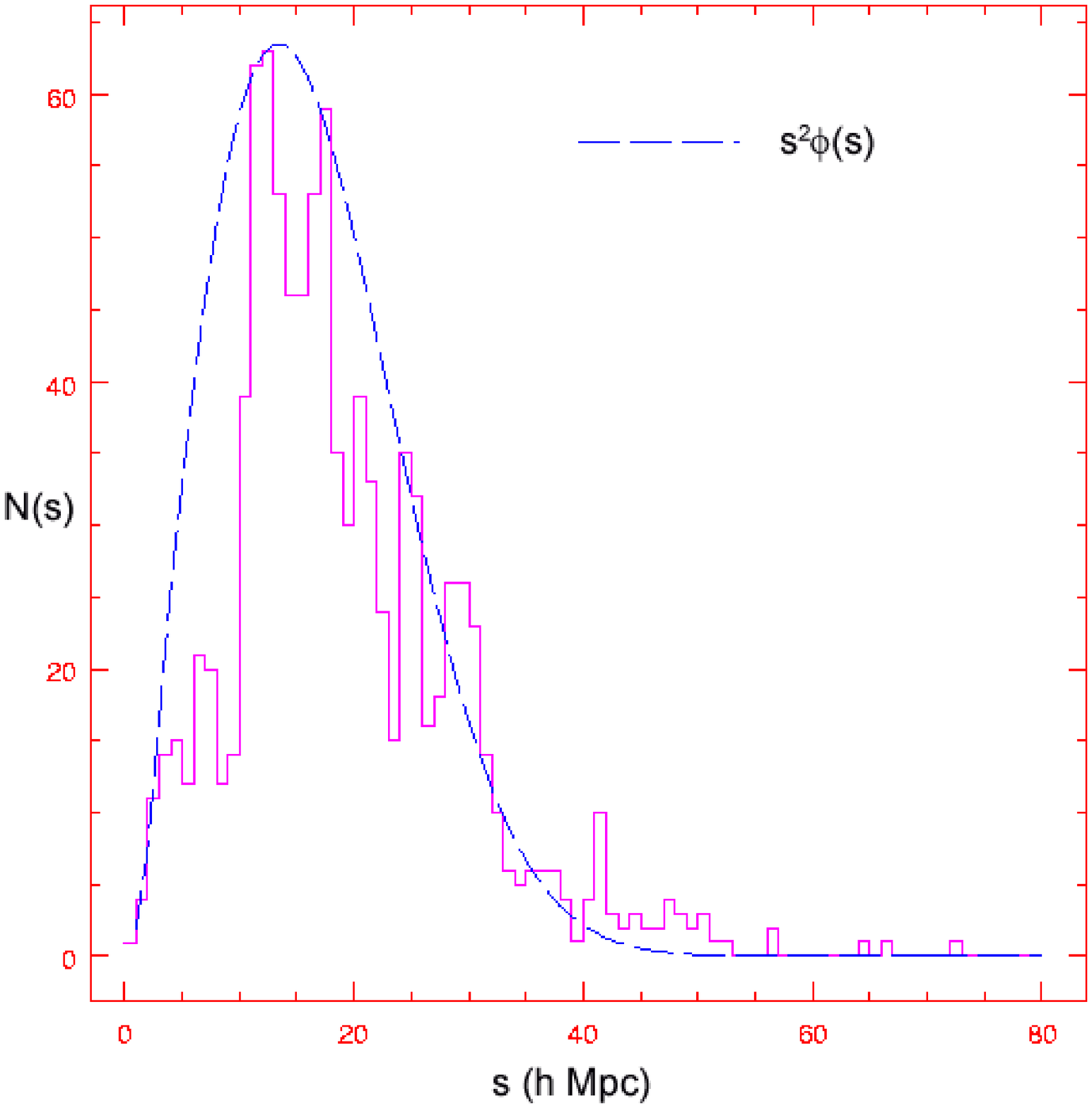}{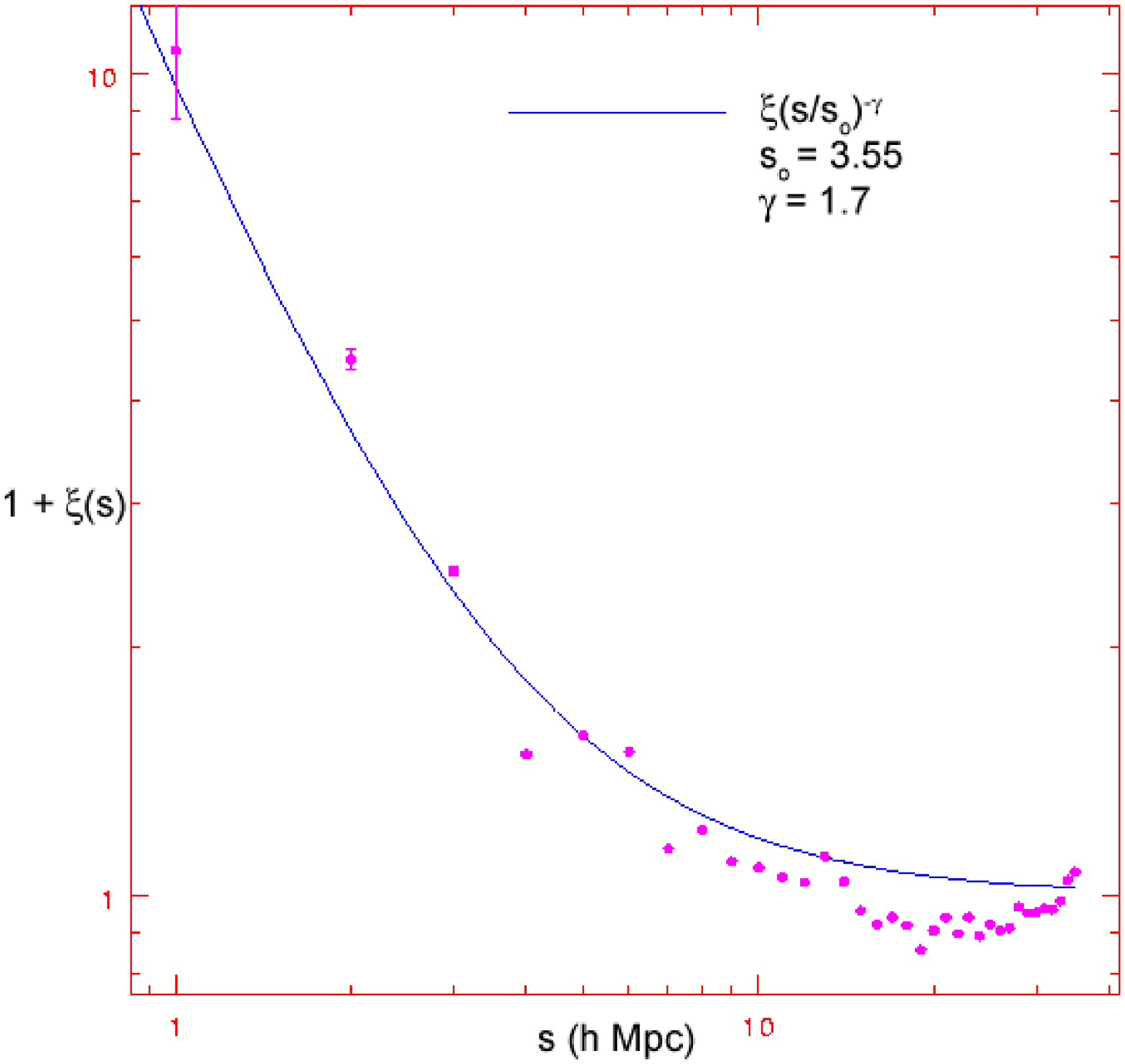}
\caption{(a) Histogram of galaxy distribution in HIPASS BGC and the selection function fitted to the histogram.; (b) The spatial CF using $\xi_{\textup{DP}}$
for the HIPASS BGC.}
\end{figure}

\noindent where $\xi_{\textup{DP}}$ refers to Davis \& Peebles estimator \cite{myref1} and $\xi_{\textup{Ham}}$ is Hamilton estimator \cite{myref3}. $DD$, $DR$ and $RR$ are the number of data-data, data-random and random-random pairs respectively. Comparing these two estimators, $\xi_{\textup{Ham}}$ is less affected by the uncertainty in the mean density, which is a second order effect. Nonetheless, even though $\xi_{\textup{DP}}$ explicitly depends on the mean density, it is the most commonly used estimator. In this analysis, both estimators have been weighted with the selection function, $\phi(s)$, and 20,000 random galaxies have been generated to match $\phi(s)$.\\
\indent Figure 1b shows a plot of $\xi(s)$ for $\xi_{\textup{DP}}$ with Poisson error bars. However this can be improved by using bootstrap method in the future work (with DEEP HIPASS catalogue). It can be fit by $\xi(s)\sim$ ($s$/$s_0$)$^{-\gamma}$ with $\gamma\sim1.7$ and $s_0\sim3.55$ $h^{-1}$ Mpc. When using $\xi_{\textup{Ham}}$, $\xi(s)$ is fit by the same form with $\gamma\sim2.19$ and $s_0\sim3.37$ $h^{-1}$ Mpc. These results definitely show less clustering than other optical selected surveys e.g. CfA redshift survey \cite{myref1} with  $\gamma\sim1.77$ and $s_0\sim5.4$ $h^{-1}$ Mpc, or Stromlo-APM with $\gamma\sim1.47$ and $s_0\sim5.9$ $h^{-1}$ Mpc \cite{myref4}.

\section{Conclusion}
Clearly, the CF from the HIPASS BGC shows less clustering than other existing surveys. However this result is expected, as HI-selected galaxies trace late-type galaxies which tend to avoid the densest region. Correspondingly, this result supports the morphology-density relation which was first described by Dressler \cite{myref5}. It will be interesting to see how the CF changes with DEEP HIPASS catalogue (available late 2001) which will contain $\sim$ 8,000 - 10,000 galaxies.


\begin{thebibliography}{}{


\bibitem{myref7} Barnes, D., et al., 2001, \mnras\ 322, 486
\bibitem{myref1} Davis, M. \& Peebles, P.J.E., 1983, \apj\ 267, 465
\bibitem{myref5} Dressler, A., 1980, \apj\ 236, 351
\bibitem{myref2} Giuricin, G., Samurovic, S., Girardi, M., Mezzetti, M. \& Marinoni, C., 2001, \apj\ 554, 872
\bibitem{myref3} Hamilton, A.J.S., 1993, \apj\ 417, 19
%\bibitem{myref4} Lapparent, V.L., Geller, M.J. \& Hucha, J.P., 1988, \apj\ 332, 44
\bibitem{myref6} Koribalski, B.S. in {\sl Gas \& Galaxy Evolution}, ASP
Conf. Series, eds J. E. Hibbard, M. P. Rupen and J. H. van Gorkom.

\bibitem{myref4} Loveday, J., Maddox, S.J., Efstathiou, G. \& Peterson, B.A., 1995, \apj\ 442, 457
%\bibitem{myref6} Peacock, J.A. \& Nicholson, D., 1991, \mnras\ 253, 307
%\bibitem{myref7} Willmer, C.N.A., da Costa, L.N. \& Pellegrini, P.S., 1998, \aj\ 115, 869
}
\end{thebibliography}
\end{document}